\documentclass[graybox]{svmult}

\usepackage{mathptmx}       % selects Times Roman as basic font
\usepackage{helvet}         % selects Helvetica as sans-serif font
\usepackage{courier}        % selects Courier as typewriter font
\usepackage{type1cm}        % activate if the above 3 fonts are
\usepackage{makeidx}         % allows index generation
\usepackage{graphicx}        % standard LaTeX graphics tool
\usepackage{gensymb}
\usepackage{amssymb}
\usepackage{amsmath}
\usepackage{bm}% bold math        
\graphicspath{{Figures/}}
\usepackage{multicol}        % used for the two-column index
\usepackage[bottom]{footmisc}% places footnotes at page bottom

\usepackage{xcolor}

\makeindex             % used for the subject index
\begin{document}

\title*{Mechanical transmission of rotation for molecule gears and solid-state gears}
\author{Huang-Hsiang Lin, Jonathan Heinze, Alexander Croy, Rafael Gutierrez and Gianaurelio Cuniberti}
\authorrunning{H-H Lin, J Heinze, A Croy, R Gutierrez, G Cuniberti}
\institute{Huang-Hsiang Lin \and Jonathan Heinze \and Alexander Croy \and Rafael Gutierrez \and Gianaurelio Cuniberti \at Institute for Materials Science and Max Bergmann Center of Biomaterials, TU Dresden, 01069 Dresden, Germany,
%\email{hhlin@nano.tu-dresden.de}}
\email{g.cuniberti@tu-dresden.de}}
% \and Christian Joachim \at GNS and MANA Satellite, CEMES-CNRS, 29 rue J. Marvig, 31055 Toulouse Cedex, France, \email{christian.joachim@cemes.fr} }
\maketitle

\abstract{
{The miniaturization of gears towards the nanoscale is a formidable task posing a variety of challenges to current fabrication technologies. In context, the understanding, via computer simulations, of the mechanisms mediating the transfer of rotational motion between nanoscale gears  can be of great help to guide the experimental designs. Based on atomistic molecular dynamics simulations in combination with a nearly rigid-body approximation, we study the  transmission of rotational motion between molecule gears and solid-state gears, respectively. For the molecule gears under continuous driving, we identify different regimes of rotational motion depending on the magnitude of the external torque. In contrast, the solid-state gears behave like ideal gears with nearly perfect transmission. Furthermore, we simulate the manipulation of the gears by a scanning-probe tip and we find that the mechanical transmission strongly depends on the center of mass distance between gears. A new regime of transmission is found for the solid-state gears.}}

%%%%%%%%%%%%%%%%%%%%%%%%%%%%%%%%%%%%%%%%%%%%%%%%%%%%%%%%%%%%%%%%%%%%%%%%%%%%%%%%%%%%%%%%%%%%%%%%%%%%%%%%%%%%%%%%%%%%%%%
\section{Introduction}\label{sec:intro}
The miniaturisation of gears down to the atomic scale, in order to transmit mechanical motion, represents a major  challenge, 
with trains of molecule gears being the ultimate target \cite{Soong2000}. To guide  ongoing experiments, it is of crucial interest to shed light on 
the microscopic features that govern the mechanics of molecule gears. In addition to  fabrication technologies based on a bottom-up approach\cite{Gisbert2019}, the production of solid-state gears
using top-down methods (e.g.\ focused ion beam\cite{JuYun2007} or electron beam\cite{Deng2011,Yang2014}) may yield a viable path towards miniaturization.

\begin{figure}[t]
    \centering
    \includegraphics[scale=0.4]{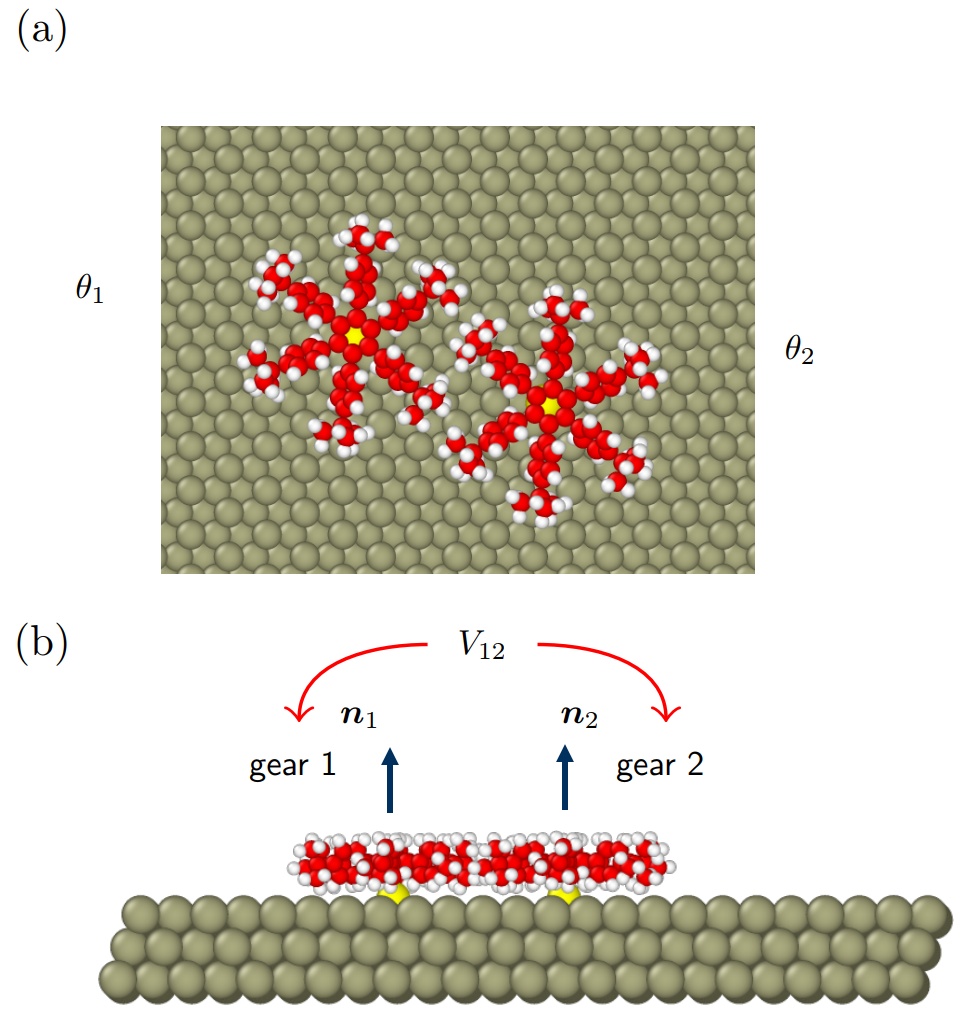}
    \caption{Schematic plots of (a) top-view (with rotation angles $\theta_1$ and $\theta_2$) and (b) side-view (with rotation axes $\boldsymbol{n}_1$ and $\boldsymbol{n}_2$) for the setup of a train of molecule gears HB-HPB mounted above Cu (yellow) atoms on top of Pb (brown) (111) surface. The interaction $V_{12}$ mediates the transmission of rotation between the gears.}
    \label{fig:scheme_STM}       % Give a unique label
\end{figure}

To manipulate molecule gears in cutting-edge experiments, typically the tip of a scanning tunneling microscope (STM) is used\cite{Gimzewski1998,Tierney2011,Ohmann2015,Perera2013,Gao2008,WeiHyo2019}. In those experiments, the molecules are deposited on a suitable substrate and moved onto nearby adatoms, whenever possible. For example, the scheme in Fig.\ \ref{fig:scheme_STM} illustrates the experimental setup reported in Ref.\ \cite{WeiHyo2019}. There, up to four hexa-\textit{t}-butyl-hexaphenylbenzene (HB-HPB) molecules were mounted on copper atoms (in yellow) on top of a lead-(111) surface (in {green}). In this situation, the molecules interact only weakly with each-other and with the substrate via van-der-Waals interactions. As it was demonstrated, by pushing one of the gears, its rotation can be transmitted to the others. 

It is interesting to compare the situation with molecule gears to the behavior of solid-state gears. For the latter, one expects perfect transmission of rotation for suitable distances between the gears. {For such gears, with mesoscopic dimensions (few nm), the number of atoms is large enough to manifest classical behavior\cite{Bombis2010}}. The main difference to the molecular case is the softness of the molecules, which influences the conditions for observing collective rotations\cite{Lin2019a}. {For the molecule gears, several atomistic calculations based on density-functional theory (DFT) and classical molecular dynamics (MD) have been carried out to investigate the transmission properties between gears. For instance, DFT has been used to study a cyclopentadienyl ring with cyano groups mounted on a manganese atom above graphene\cite{Hove2018}, as well as PF$_3$ molecules on a Copper(111) surface\cite{Hove2018a} {(see also the chapter by Srivastava \textit{et al.} in this volume)}. In particular, the influence of the flexibility of gears and the slippage between gears has been investigated. MD simulations have been performed for carbon nanotube, fullerene-based and molecule gears\cite{Han1997, Robertson1994, Lin2019a}. But at the moment, a direct comparison of different gears in terms of the mechanical transmission between them is still missing.} {Therefore, a systematic analysis for different type of gears, separation distance and external driving is of particular interest.}

In this chapter, we focus on the mechanical transmission of motion in both molecule-based gears and nanoscale solid-state gears, and investigate the conditions under which  collective rotation is possible. In particular, we compare the results for molecule gears based on HB-HPB with those for solid-state gears (diamond) using the same {model for the substrate and the same temperature}. We use all-atom molecular dynamics (MD) simulations to investigate the problem, since it  allows  to reach  relevant timescales of about $100$ps to $1$ns, even for solid-state gears. The simulations  also yield trajectories longer than the surface relaxation time, which is on the order of few picoseconds\cite{Persson1996}. The trajectories are analyzed using a nearly-rigid body approximation (NRBA), which enables a separation of the rigid-body motion and the internal deformation of the gears\cite{Lin2019a}.

The chapter is organized as follows: in Sec.\,\ref{sec:modelling}, we introduce the nearly rigid-body approximation (Sec.\,\ref{subsec:NRBA}) and the details of the MD simulations (Sec.\,\ref{subsec:MD}) for molecule and solid-state gears. In Sec.\,\ref{sec:Result}, we show and discuss the results for a train of molecule and solid-state gears driven by an external torque (Sec.\,\ref{subsec:external_torque}) and under tip manipulation (Sec.\,\ref{subsec:tip_manipulation}). Finally, in Sec.\,\ref{sec:conclusion}, we summarize our results and provide  an outlook.

%%%%%%%%%%%%%%%%%%%%%%%%%%%%%%%%%%%%%%%%%%%%%%%%%%%%%%%%%%%%%%%%%%%%%%%%%%%%%%%%%%%%%%%%%%%%%%%%%%%%%%%%%%%%%%%%%%%%%%%
\section{Modelling}\label{sec:modelling}
In this section, we will introduce the NRBA to define the orientation vectors of individual gears and describe the setup of the MD simulations for a train of molecule gears and solid-state gears.
\begin{figure}[t]
    \centering
    \includegraphics[scale=0.9]{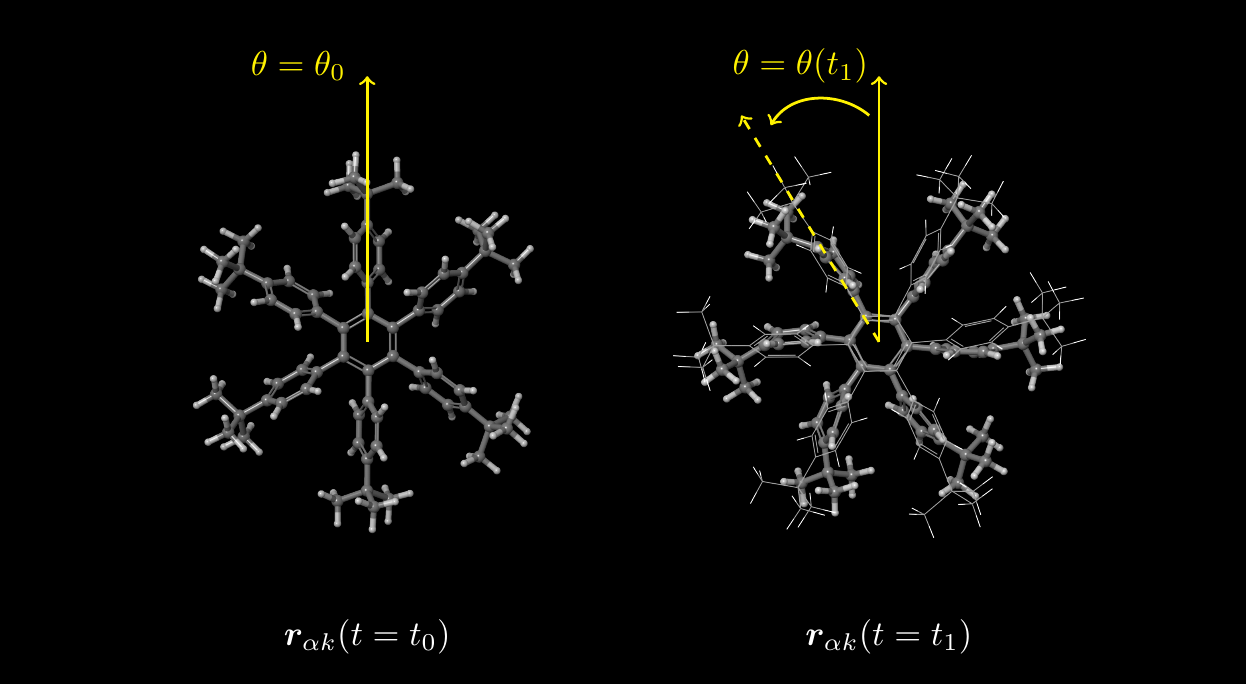}
    \caption{Demonstration of the nearly rigid-body approximation. On the left, the initial molecular geometry at $t=t_0$ is set as a reference frame with orientation $\theta=\theta_0$; on the right, the deformed structure at $t=t_1$ is mapped to the best-fitting rigid-body transformation (with thinner sticks) from the reference frame with the uniquely-defined orientation $\theta=\theta(t_1)$.}
    \label{fig:NRBA}       % Give a unique label
\end{figure}
\subsection{Nearly Rigid-Body Approximation}\label{subsec:NRBA}
In order to define the orientation-vector of the molecule and solid-state gears, we use the NRBA as introduced in Ref.\ \cite{Lin2019a}. First, we consider a train of gears as shown in Fig.\ \ref{fig:scheme_STM}. For each gear, we define an appropriate reference structure represented by a set of Cartesian coordinates $\{{\bm{r}^0_{\alpha k}}\}$, where $\alpha$ denotes the gear index and $k$ is running over all atoms in the $\alpha^{th}$ gear. For instance, we can choose the initial frame $\{{\bm{r}_{\alpha k}(t_0)}\}$ of the MD simulation
which corresponds to the optimized geometry of the molecule. Secondly, the gear geometry at a later time $t$ is given by \{${\bm{r}_{\alpha k}(t)}$\} (the structure on the right in Fig.\ \ref{fig:NRBA}). Next, we assume that the deformation of the gear during the simulation is sufficiently small, so that we can always find a unique set of rotational axes $\bm{n}_{\alpha}$ and angles $\theta_{\alpha}$. Those define the best-fitting rigid-body rotation transformation $\bm{R}^{\alpha}(\theta_{\alpha },\bm{n}_{\alpha})$ of the reference structure (the thinner structure on the right panel in Fig.\ \ref{fig:NRBA}) to the current structure. {At the same time, the deviation from the best-fitting transformation is defined as deformation.} 

To be specific, the {\textit{deformation}} for a given structure $\bm{r}_{\alpha k}$ in the NRBA is defined as:
\begin{equation}\label{eq:loc_error}
    \bm{\epsilon}_{\alpha k} =  \bm{R}^{\alpha}(\theta_{\alpha },\bm{n}_{\alpha})\bm{r}_{\alpha k}^{0}-\bm{r}_{\alpha k}\;. 
\end{equation}
The weighted sum of squared deformation for all atoms of the $\alpha^{th}$ gear is given by:
\begin{equation}\label{eq:tot_error}
    \epsilon^{\alpha}_{tot} = \sum_{k}^{}w_{\alpha k}|\bm{\epsilon}_{\alpha k}|^{2}\;,
\end{equation}
where the positive weight is taken to be $w_{\alpha k} = m_{\alpha k} / \sum_{k}m_{\alpha k}$, i.e.\ the ratio between the individual mass $m_{\alpha k}$ of the $k^{th}$ atom and the total mass $\sum_{k}m_{\alpha k}$ of gear $\alpha$. {This implies a larger contribution to the total deformation for heavier atoms}.
 
Technically, the best-fitting transform $\bm{R}^{\alpha}(\theta_{\alpha },\bm{n}_{\alpha})$ can be found by using quaternions \cite{Kneller1991}. The latter are defined by four numbers, $\underline{q}^{\alpha}=(q_0^{\alpha},q_1^{\alpha},q_2^{\alpha},q_3^{\alpha})$ (for simplicity, we suppress the $\alpha$ index  in what follows). The rotation matrix is then related to the quaternions via:
\begin{equation}\label{eq:R_matrix}
    \bm{R}^{\alpha }(q) =
    \begin{pmatrix}
    q_0^{2}+q_1^{2}-q_2^{2}-q_3^{2}  &&  2(-q_0q_3+q_1q_2)  &&  2(q_0q_2+q_1q_3)\\
    2(q_0q_3+q_1q_2)  &&  q_0^{2}+q_2^{2}-q_1^{2}-q_3^{2}  &&  2(-q_0q_1+q_2q_3)\\
    2(-q_0q_2+q_1q_3)  &&  2(q_0q_1+q_2q_3)  &&  q_0^{2}+q_3^{2}-q_1^{2}-q_2^{2}
    \end{pmatrix}\;.
\end{equation}
Accordingly, the quaternion components are related to the rotation axes $\bm{n}_\alpha=(n^{\alpha}_{x},n^{\alpha}_{y},n^{\alpha}_{z})$ and the rotation angle $\theta_{\alpha}$ by:
\begin{eqnarray}
     q_0 &=& \cos(\theta_{\alpha }/2)\;,\nonumber\\
     q_1 &=& \sin(\theta_{\alpha }/2)n^{\alpha}_{x}\;,\nonumber\\
     q_2 &=& \sin(\theta_{\alpha }/2)n^{\alpha}_{y}\;,\nonumber\\
     q_3 &=& \sin(\theta_{\alpha }/2)n^{\alpha}_{z}\;.
\end{eqnarray}

{In order to obtain the best rigid-body transform $\bm{R}^{\alpha }(q)$, we 
insert Eq.\ \eqref{eq:R_matrix} into Eq.\ \eqref{eq:loc_error}, and subsequently minimize Eq.\ \eqref{eq:tot_error} with respect to  $\underline{q}^{\alpha}$, and subject to the normalization condition $\underline{q}^{\alpha}\cdot \underline{q}^{\alpha}=1$. Equivalently, the quaternion $\underline{q}^{\alpha}$ can be obtained by minimizing the following function via the method of Lagrange multipliers:
\begin{equation}
     f(\underline{q}^{\alpha},\lambda^{\alpha})=\epsilon^{\alpha}_{tot}(\underline{q}^{\alpha})-\lambda^{\alpha}(\underline{q}^{\alpha}\cdot \underline{q}^{\alpha}-1)\;.
\end{equation}
This results in the eigenvalue problem:
\begin{equation}\label{eq:eigenvalue}
     \bm{M}_{\alpha}\underline{q}^{\alpha}=\lambda^{\alpha} \underline{q}^{\alpha}\qquad\text{with}\qquad
     \underline{q}^{\alpha}\cdot \underline{q}^{\alpha}=1 \;,
\end{equation}
where the matrices $\bm{M}_{\alpha}$} can be shown to depend directly on $\bm{r}_{\alpha k}=(x_{\alpha k},y_{\alpha k},z_{\alpha k})$ and $\bm{r}^0_{\alpha k}=(x^0_{\alpha k},y^0_{\alpha k},z^0_{\alpha k})$ \cite{Kneller1991}.
More explicitly,
\begin{equation}
     \bm{M}_{\alpha} = \sum_{k}w_{\alpha k}\bm{M}_{\alpha k}\;
\end{equation}
with the independent components of the symmetric matrix $\bm{M}_{\alpha k}$ given by:
\begin{eqnarray}
     (\bm{M}_{\alpha k})_{11} &=& x^2_{\alpha k}+y^2_{\alpha k}+z^2_{\alpha k}+(x^0_{\alpha k})^2+(y^0_{\alpha k})^2+(z^0_{\alpha k})^2-2(x_{\alpha k}x^0_{\alpha k}+y_{\alpha k}y^0_{\alpha k}+z_{\alpha k}z^0_{\alpha k})\;,\nonumber\\
    (\bm{M}_{\alpha k})_{12} &=& 2(y_{\alpha k}z^0_{\alpha k}-z_{\alpha k}y^0_{\alpha k})\;,\nonumber\\
     (\bm{M}_{\alpha k})_{13} &=& 2(-x_{\alpha k}z^0_{\alpha k}+z_{\alpha k}x^0_{\alpha k})\;,\nonumber\\
     (\bm{M}_{\alpha k})_{14} &=& 2(x_{\alpha k}y^0_{\alpha k}-y_{\alpha k}x^0_{\alpha k})\;,\nonumber\\
     (\bm{M}_{\alpha k})_{22} &=& x^2_{\alpha k}+y^2_{\alpha k}+z^2_{\alpha k}+(x^0_{\alpha k})^2+(y^0_{\alpha k})^2+(z^0_{\alpha k})^2-2(x_{\alpha k}x^0_{\alpha k}-y_{\alpha k}y^0_{\alpha k}-z_{\alpha k}z^0_{\alpha k})\;,\nonumber\\
     (\bm{M}_{\alpha k})_{23} &=& -2(x_{\alpha k}y^0_{\alpha k}+y_{\alpha k}x^0_{\alpha k})\;,\nonumber \\
     (\bm{M}_{\alpha k})_{24} &=& -2(x_{\alpha k}z^0_{\alpha k}+z_{\alpha k}x^0_{\alpha k})\;,\nonumber\\
    (\bm{M}_{\alpha k})_{33} &=& x^2_{\alpha k}+y^2_{\alpha k}+z^2_{\alpha k}+(x^0_{\alpha k})^2+(y^0_{\alpha k})^2+(z^0_{\alpha k})^2+2(x_{\alpha k}x^0_{\alpha k}-y_{\alpha k}y^0_{\alpha k}+z_{\alpha k}z^0_{\alpha k})\;,\nonumber\\
    (\bm{M}_{\alpha k})_{34} &=& -2(y_{\alpha k}z^0_{\alpha k}+z_{\alpha k}y^0_{\alpha k})\;,\nonumber\\
    (\bm{M}_{\alpha k})_{44} &=& x^2_{\alpha k}+y^2_{\alpha k}+z^2_{\alpha k}+(x^0_{\alpha k})^2+(y^0_{\alpha k})^2+(z^0_{\alpha k})^2+2(x_{\alpha k}x^0_{\alpha k}+y_{\alpha k}y^0_{\alpha k}-z_{\alpha k}z^0_{\alpha k})\nonumber\;.
\end{eqnarray}
{Finally, the quaternion which is minimizing the deformation is given by the eigenvector of Eq.\ \eqref{eq:eigenvalue} with the smallest eigenvalue. The degree of deformation is directly given by the corresponding eigenvalue.}

In summary, the NRBA allows us to extract the rigid-body transformation and the deformation connecting two arbitrary configurations as illustrated in Fig.\ \ref{fig:NRBA}.
\begin{figure}[t]
    \centering
    \includegraphics[scale=0.6]{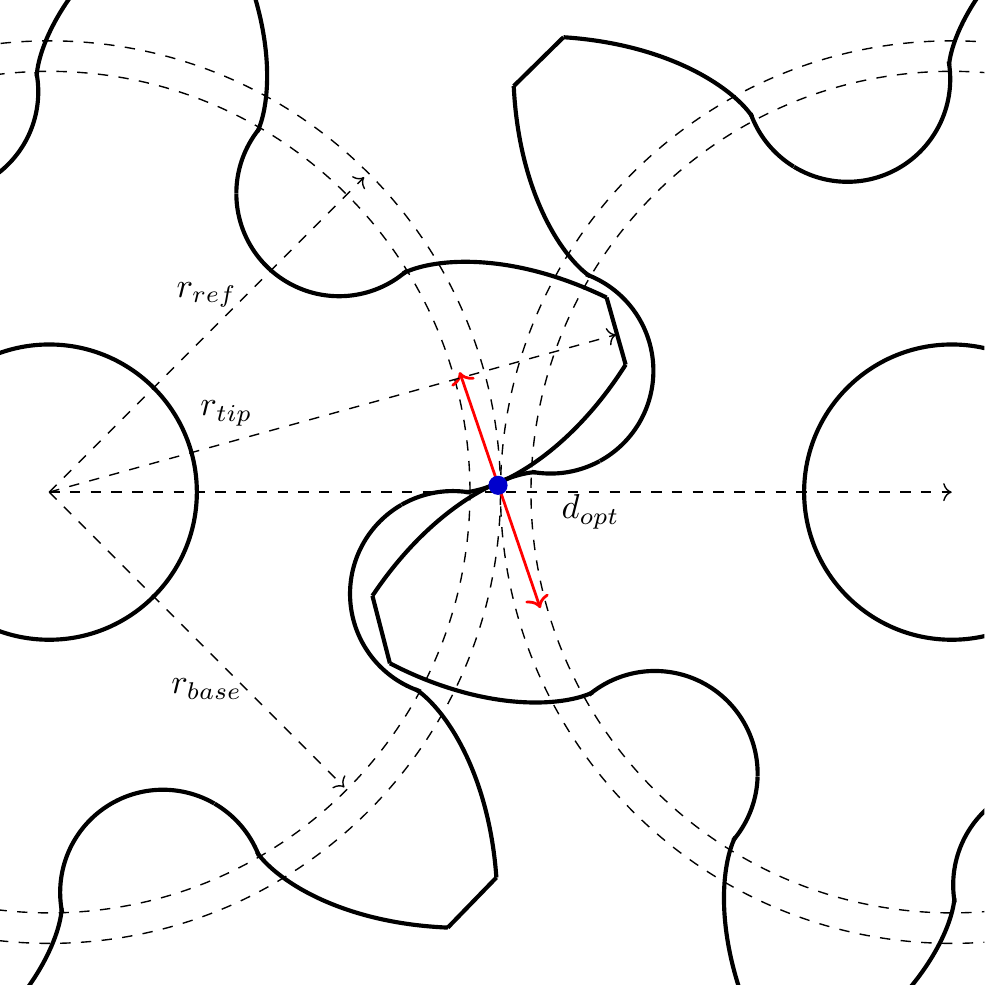}
    \caption{Demonstration of two interlocked solid-state gears. The single-point contact (pressure point) is marked by a blue dot in the center. As the gears rotate, the pressure point moves along the line of action (red line) which will stay tangent to the base circles with radius $r_{base}$ at all times for optimal transfer of angular momentum. Since we use multiple gears with equal dimensions, the optimal distance is given by the center of mass distance between two gears $d_{opt} = 2 \cdot r_{ref}$. The gear size $r_{tip}$ is defined as the distance between the center of mass and the gear tip.}
    \label{fig:meshing}       % Give a unique label
\end{figure}

\subsection{Solid-state gear meshing}
\label{subsec:meshing}
In order to create the solid state gears, we follow a general algorithm for creating involute spur gears \cite{Norton2010}. We  then use the Open Visualization Tool (OVITO) \cite{Stukowski2010} to cut the gears from a bulk diamond crystal. {To be specific, the typical structure to define gears is shown in Fig.}\ \ref{fig:meshing}. The figure shows the contact between two ideal involute gears;  they touch each other at a single point (pressure point) marked by the blue dot in the center. As the gears rotate, the pressure point moves along the line of action (red line) which will stay tangent to the base circles with radius $r_{base}$ at all times for optimal transfer of angular momentum. Since we use multiple gears with equal dimensions, the optimal distance is given by the center of mass distance between two gears $d_{opt} = 2 \cdot r_{ref}$. For later discussion, we define the gear size by $r_{tip}$, which is the distance between the center of mass and a gear tip.

\subsection{Molecular Dynamics}
\label{subsec:MD}
Since we focus on the rotational transmission between gears, {we model our problem based on the following general assumptions for both solid-state gears and molecule gears}: 
\begin{enumerate}
    \item The gears are weakly coupled to the surface. Therefore,  charge transfer effects between the two systems can be neglected and the specific atomic position on the surface is not relevant.
    \item The gears are well anchored, which can  be mimicked by fixing the centers of mass.
    \item The gears are initially in thermal equilibrium with the surface.
\end{enumerate}
To be specific, we use the Large-scale Atomic/Molecular Massively Parallel Simulator (LAMMPS) \cite{Plimpton1995} for implementing the MD simulations. Based on the first assumption above, we use an artificial van-der-Waals surface, which interacts with the molecules via a 9-3 Lennard-Jones-Potential:
\begin{equation}
    V_{LJ}(r) = \epsilon\left[\frac{2}{15}\left(\frac{\sigma}{r}\right)^{9}-\left(\frac{\sigma}{r}\right)^{3}\right]\;.
\end{equation}
Here, we use $\epsilon = 0.1$ eV, $\sigma = 5$ \AA\, and an initial distance of $5$ \AA{} between the surface and the gears. Then, according to the third assumption, we use a Langevin thermostat with the relaxation time $\tau = 1$ ps\cite{Persson1996}. For the interatomic potentials describing the molecular gear and diamond-based solid-state gear, respectively, we use the adaptive intermolecular reactive empirical bond order (AIREBO) potential \cite{Stuart2013}, which is suitable for simulations of hydrocarbons. {In all simulations, we set the temperature to $T=10$ K, mimicking  the typical conditions of a low-temperature STM experiment. Before running the simulation, the total system undergoes a geometry optimization by the conjugate gradient method built into LAMMPS.}
\begin{figure}[t]
    \centering
    \includegraphics[width=\textwidth]{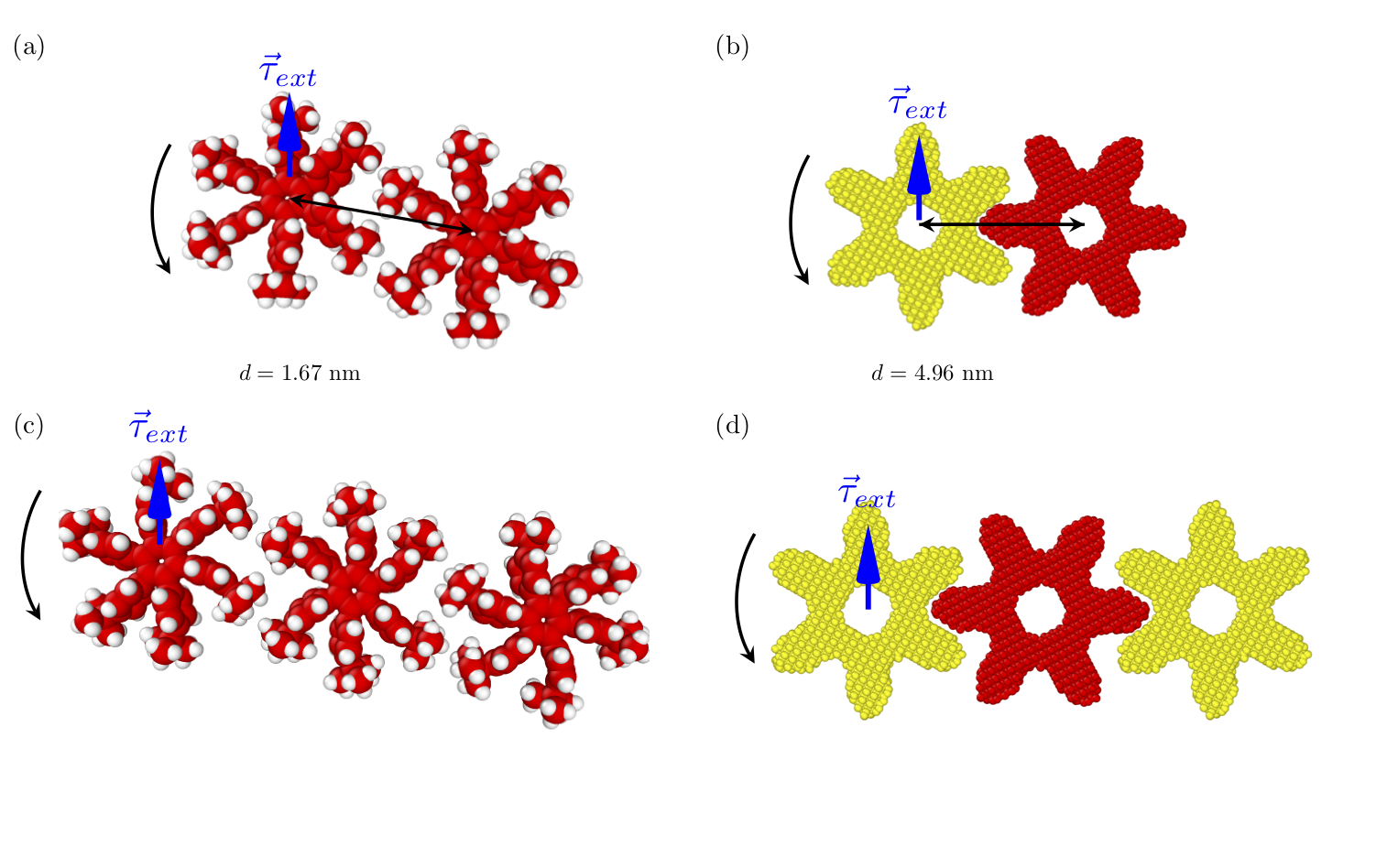}
    \caption{Schematic plots for driving gears with external torque $\vec{\tau}_{ext}$ (orientations are indicated by blue arrow) applied to the first gear on the left. (a) Two and (c) three molecule gears with center of mass distance $d=1.67$ nm; (b) two and (d) three diamond-based solid-state gears with distance $d=5$ nm.}
    \label{fig:gear_scheme}       % Give a unique label
\end{figure}
% 

%%%%%%%%%%%%%%%%%%%%%%%%%%%%%%%%%%%%%%%%%%%%%%%%%%%%%%%%%%%%%%%%%%%%%%%%%%%%%%%%%%%%%%%%%%%%%%%%%%%%%%%%%%%%%%%%%%%%%%%
\section{Results and Discussion}\label{sec:Result}
In this section, we treat two different methods to rotate gears by either applying an external torque to one of the gears in a train, or by moving a gear via a realistic tip manipulation. We compare the locking coefficients and transmission coefficients, which provide a measure for the transmission quality and which will be defined below, for both a train of molecule gears and solid-state gears (with radius $r=3$ nm {and 4932 atoms}). In the following, we  discuss the two methods in more detail.
\begin{figure}[t]
    \centering
    \includegraphics[width=\textwidth]{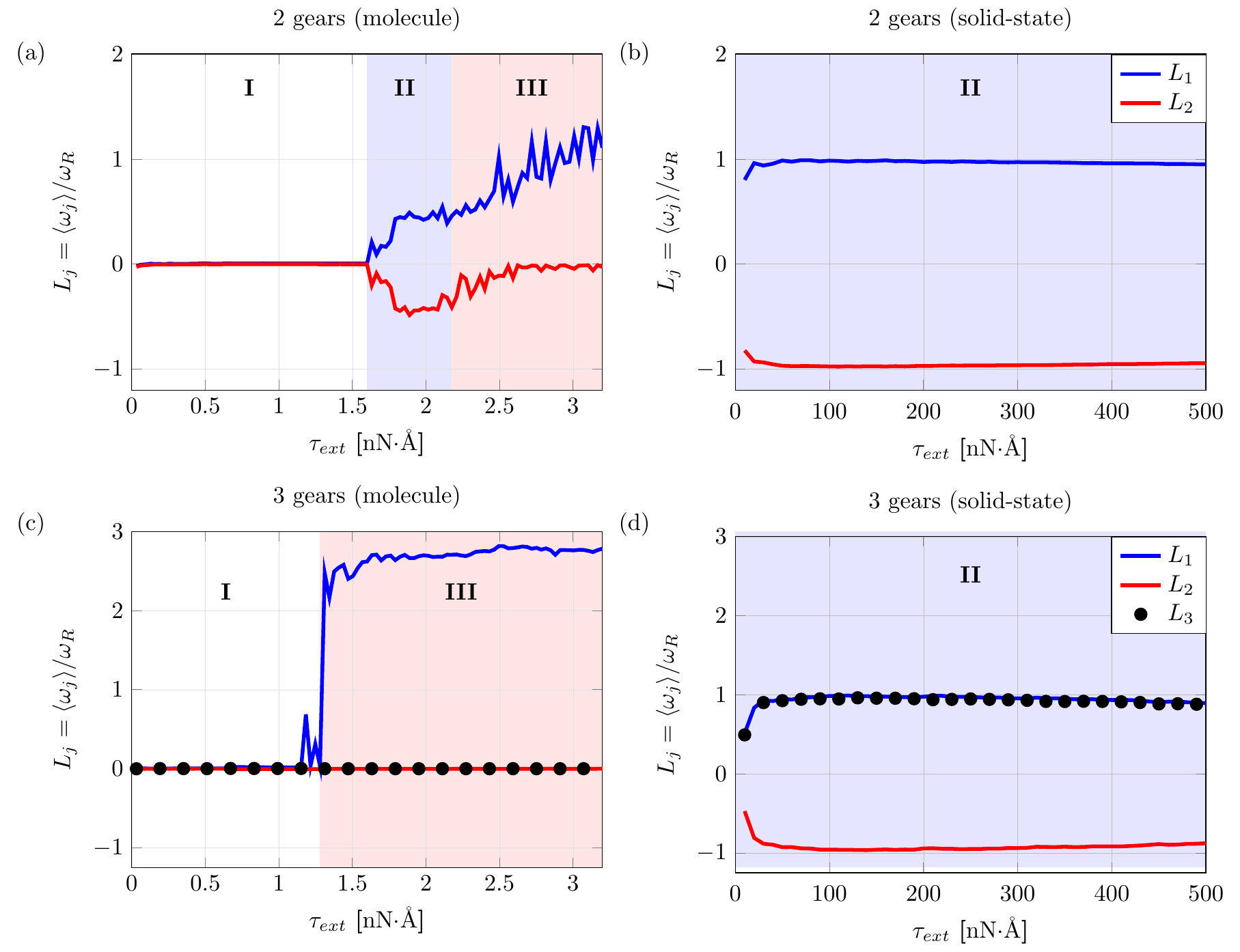}
    \caption{The locking coefficient $L_j=\langle \omega_j\rangle/\omega_{R}$ for (a) two and (c) three molecule gears with center of mass distance $d=1.67$ nm; and (b) two and (d) three solid-state gears, where $\langle \omega_j\rangle$ denotes the average angular velocity of $j^{th}$ gear and $\omega_R$ represents the terminal angular velocity of rigid-body gear. The region I (white), II (blue) and III (red) indicate the underdriving ($\mid L_i\mid\approx\mid L_j\mid\approx 0$), driving ($0<\mid L_i\mid \approx \mid L_j\mid \leqslant 1$) and overdriving ($0\approx \mid L_i \mid\approx \mid L_j\mid\ll L_1$)  regimes, respectively.}
    \label{fig:Locking_coeff}      
\end{figure}
\begin{figure}[t]
    \centering
    \includegraphics[width=\textwidth]{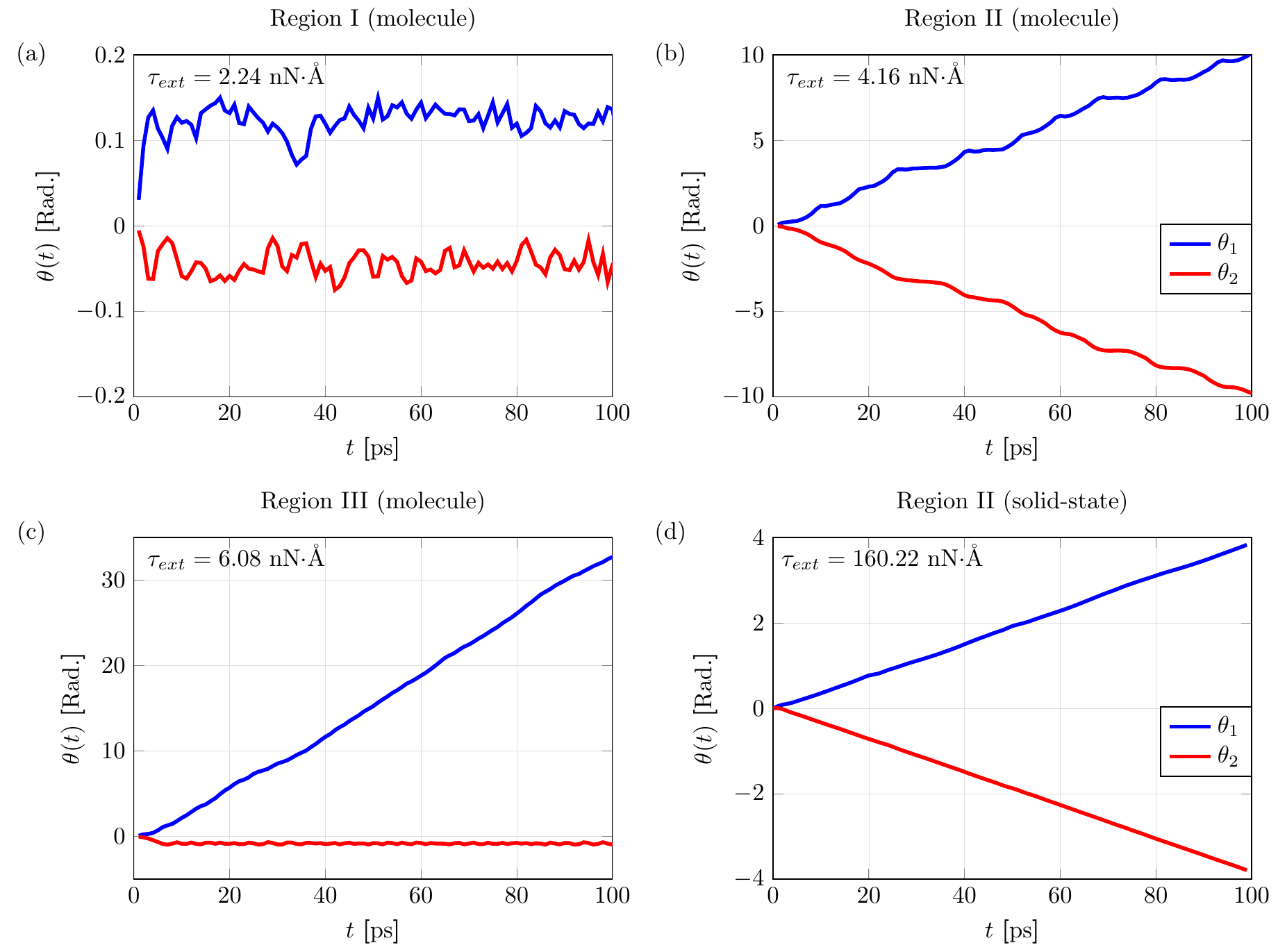}
    \caption{The trajectories within 100 ps for (a) region I (underdriving) with external torque $\tau_{ext}=2.24$ nN$\cdot$\AA\,, (b) II (driving) with $\tau_{ext}=4.16$ nN$\cdot$\AA\, and (c) III (overdriving) with $\tau_{ext}=6.08$ nN$\cdot$\AA\, of two molecule gears. (d) The trajectories for two solid-state gears with $\tau_{ext}=160.22$ nN$\cdot$\AA\ (only region II exists).}
    \label{fig:trajectories}       % Give a unique label
\end{figure}
 
\subsection{Rotation driven by an external torque}\label{subsec:external_torque}
First, we consider the scenario shown in Fig.\ \ref{fig:gear_scheme}, where  we apply an external torque $\tau_{ext}$ (with the blue arrow pointing to $+z$ direction) to the first gear on the left, which would in turn drive the neighboring gear counterclockwise. 
Moreover, depending on molecule gear or solid-state gear, one has to decide the center of mass distance $d$ between gears. For{HB-HPB} gears, the maximal distance with interlocking is shown to be between $1.67$ nm and $1.74$ nm. Here we use $d=1.67$ nm. For solid-state gears, since we use the standard spur gear, the optimal distance for the $3$ nm gear is $4.5$ nm. However, in reality, the atoms cannot be arbitrarily close due to the strong repulsion, therefore we adjust the distance to $5$ nm.

{Once the distances} are specified, we run the MD simulations to study the response of the gears to the external torque. The results are shown in Fig.\ \ref{fig:Locking_coeff}. In order to characterize the transmission of motion across gears, we define the \textit{locking coefficient} as follows:
\begin{equation}
    L_j=\frac{\langle \omega_j \rangle}{\omega_R}\;.
\end{equation}
Here, $\langle \omega_j\rangle$ denotes the average angular velocity of the $j^{th}$ gear and $\omega_R$ represents the terminal angular velocity of perfectly interlocked rigid-body gears in a train with $N$ gears. The terminal angular velocity is given by:
\begin{equation}
    \omega_R=\frac{\tau_{ext}}{N\gamma}=\frac{\tau_{ext}\tau}{N I}\;,
\end{equation}
where $\gamma$ is the damping coefficient given by $\gamma=I/\tau$ with $N=2$ or $3$, the moment of inertia is $I=2.13\times 10^{-41}$ kg$\cdot$m$^2${\cite{Lin2019a}} and the relaxation time is set to $\tau=1$ ps\cite{Persson1996}. 

{The locking coefficient provides a measure for the ability to transfer rotations between gears. For perfectly interlocked gears, the coefficient is equal to $\pm 1$.} In Fig.\ \ref{fig:Locking_coeff} (a), we show results for a MD simulation of two molecule gears (as in in Fig.\ \ref{fig:gear_scheme} (a)) within $100$ ps. We compute the dependence of the locking coefficients to the external torque, which is ramped up from $0$ to $3.2$ nN$\cdot$\AA{}. One can see that there are three different regions of motion for gears (highlighted in white, blue and red){\cite{Lin2019a}}. 

\paragraph{Region I:} 
\begin{equation}
    \mid L_1\mid\approx\mid L_2\mid\approx 0\;.
\end{equation}
For $0<\tau_{ext}<1.6$ nN$\cdot$\AA, both locking coefficients are  vanishing, meaning that the gears barely rotate. The corresponding trajectories $\theta_1$ and $\theta_2$ with $\tau_{ext}=2.24$ nN$\cdot$\AA\, are shown in Fig.\ \ref{fig:trajectories} (a), which correspond to typical Brownian rotations at finite temperature. In this case, we say that the two gears are in the \textit{underdriving phase}.

\paragraph{Region II:} 
\begin{equation}
    0<\mid L_1\mid\approx\mid L_2\mid \leqslant 1\;.
\end{equation}
For $1.6<\tau_{ext}<2.2$ nN$\cdot$\AA, the locking coefficients are approximately opposite to each other, which means that the gears are interlocked. The corresponding trajectories $\theta_1$ and $\theta_2$ with $\tau_{ext}=4.16$ nN$\cdot$\AA\, are shown in Fig.\ \ref{fig:trajectories} (b), which represent the pattern of step-by-step collective rotation. We denote this case as the \textit{driving phase}. One can see that, for this type of molecule gear, the locking coefficient $\mid L_j\mid$ is around 0.5 in the driving phase, which indicates that the gears are rather soft and some energy is dissipated into the internal degrees of freedom in the form of deformations.

\paragraph{Region III:} 
\begin{equation}
 0\approx\mid L_2\mid \ll  L_1 \leqslant N\;.
\end{equation}
For $\tau_{ext}>2.2$ nN$\cdot$\AA, the locking coefficient $L_1$ is much larger than all the others, so that only the first gear rotates. The corresponding trajectories $\theta_1$ and $\theta_2$ with $\tau_{ext}=6.08$ nN$\cdot$\AA\, are shown in Fig.\ \ref{fig:trajectories} (c), and represent the pattern of a single gear rotation. In this case, we say that the two gears are in the \textit{overdriving phase}. Note that in the overdriving phase $L_1$ may be larger than one but it has to be bounded by the  terminal velocity of free single gear rotation, namely:
\begin{equation}
    L_1=\frac{\omega_1}{\omega_R}\leqslant \frac{\tau_{ext}/\gamma}{\tau_{ext}/(N\gamma)}=N \;.
\end{equation}

We can do a similar  same analysis for three molecule gears as shown in Fig.\ \ref{fig:gear_scheme} (c). In this case, one  immediately sees that there are only two regions (I and III): underdriving phase for $0<\tau_{ext}<1.2$ nN$\cdot$\AA\,and overdriving phase for $\tau_{ext}> 1.2$ nN$\cdot$\AA\,. This implies that no matter how hard the first gear is driven, it is not possible to have a collective rotation due to the softness of the molecules.
\begin{figure}[t]
    \centering
    \includegraphics[width=\textwidth]{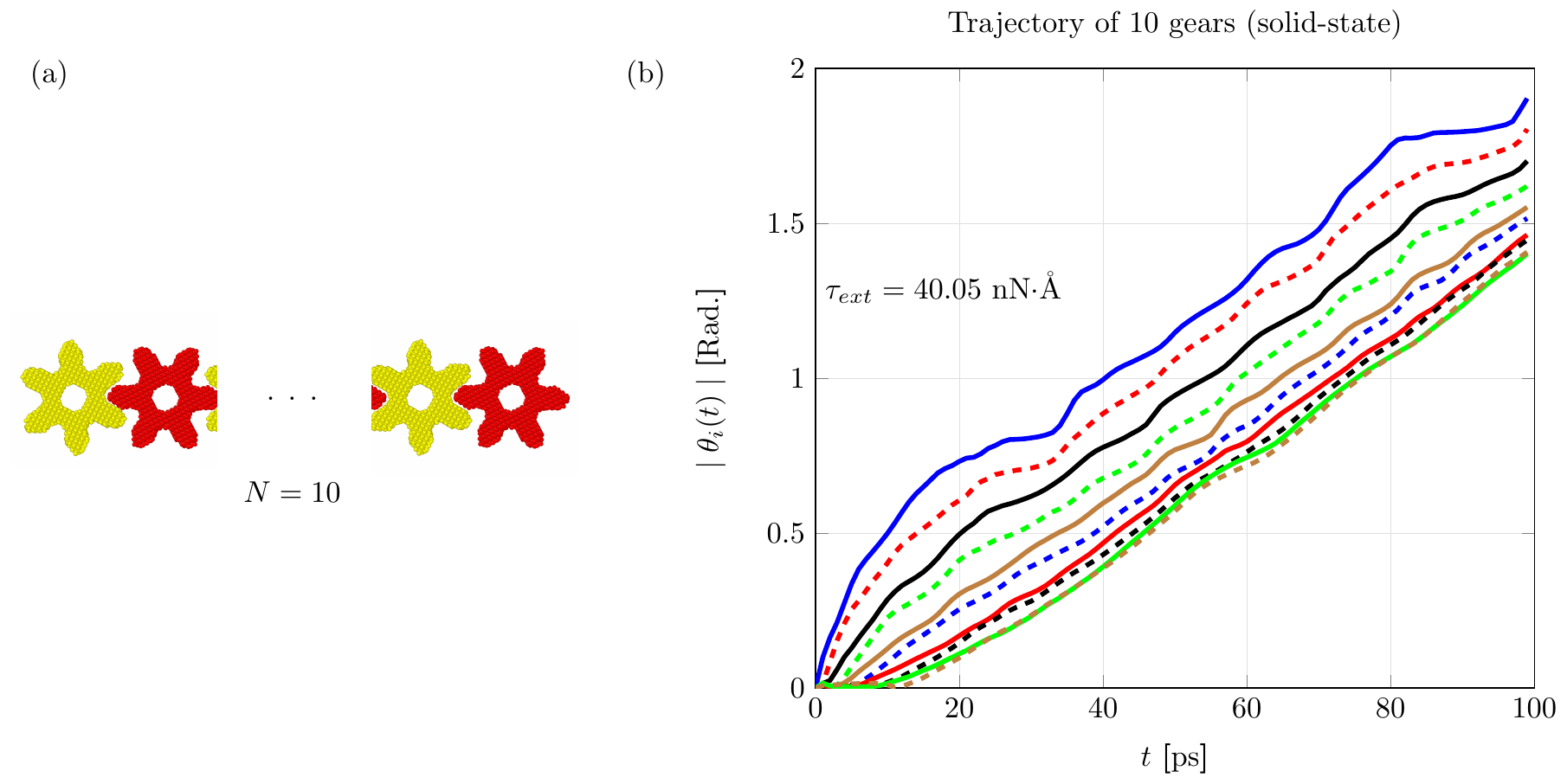}
    \caption{(a) Schematic configuration of a gear train consisting of $N=10$ gears driven by an external torque $\tau_{ext}=40.05$ nN$\cdot$\AA$ $. (b) The trajectories (solid (dashed) lines represent counterclockwise (clockwise)) in \textit{driving phase}.}
% One can see that the propagation of rotation is delayed. The delay between the first ($i=1$, blue line) and the last gear ($i=10$, dashed brown line) is not constant but ranges roughly between $15$ ps $\leq \Delta t\leq 50$ ps
    \label{fig:ten_gears}       % Give a unique label
\end{figure}
 
For comparison, we move on to the solid-state gears as shown in Fig.\ \ref{fig:gear_scheme} (b) and (d). Since the gear is based on diamond, which is the hardest material, one expects a rather stiff or rigid behavior. As one can see in Fig.\ \ref{fig:Locking_coeff} (b) and (d), only region-II behavior appears, so that gears are always in the driving phase. On the other hand, the locking coefficients $L_j$ are close to one,  indicating a rigid-body interlocked rotation. This is also consistent with the trajectories obtained for $\tau_{ext}=160.22$ nN$\cdot$\AA\, in Fig.\ \ref{fig:trajectories} (d).

Since the solid-state gears are rather rigid, one can see that the collective rotation happens even in the ten gears case as shown in Fig.\ \ref{fig:ten_gears}. Besides the collective rotation, there is a delay time between the gear response. For instance, the total propagation time from the first gear to the last one lies approximately  between $15$ ps $\leq \Delta t\leq 50$ ps.

\subsection{Rotation via tip manipulation}\label{subsec:tip_manipulation}
\begin{figure}[t]
    \centering
    \includegraphics[width=\textwidth]{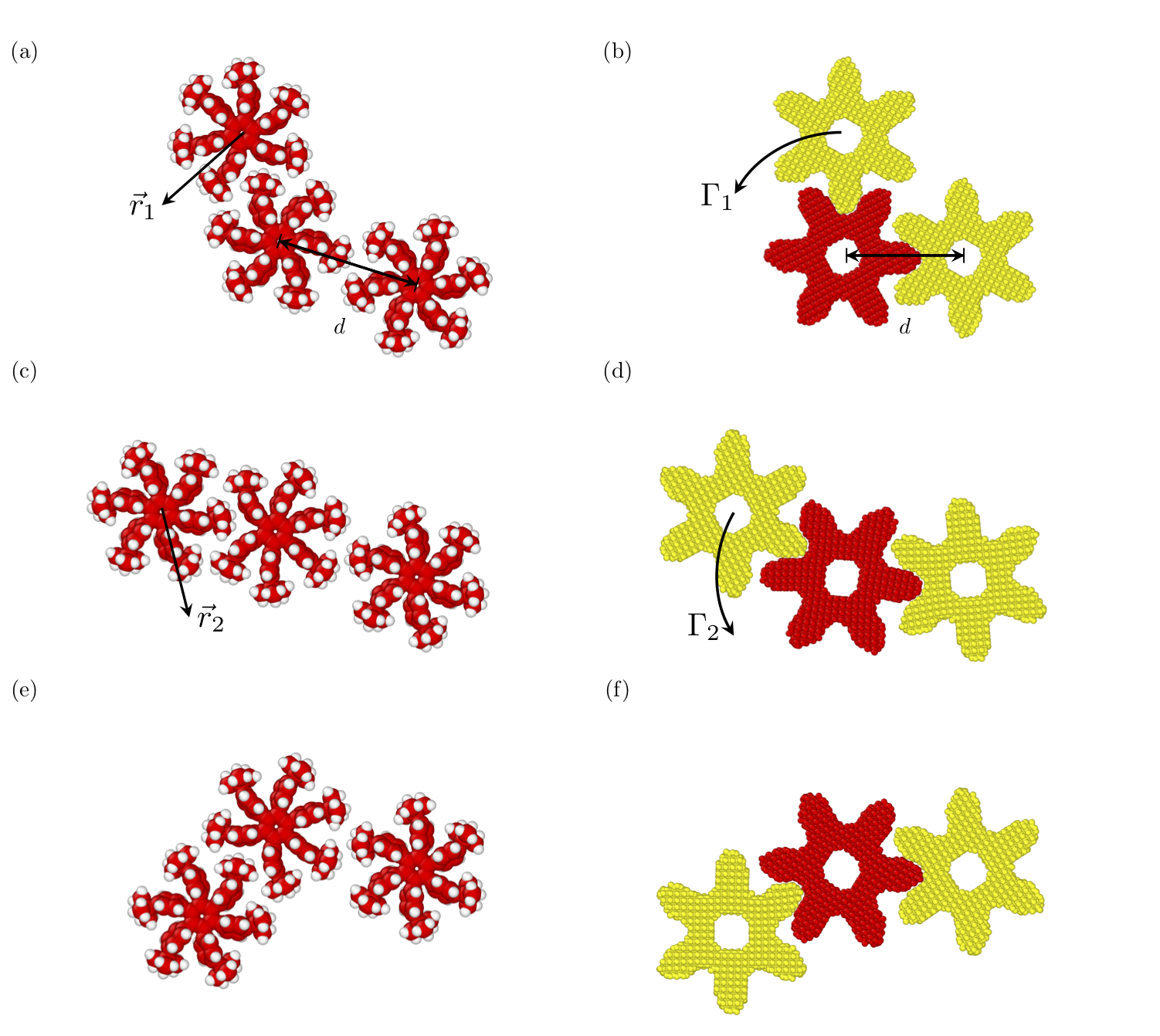}
    \caption{Schematic plots for manipulating the first gear on the left for both molecule gears and solid-state gears with constant speed $v = 3$ m/s. (a), (c) and (e) show three different conformations during the \textit{linear} two-step manipulation (with manipulation vectors $\vec{r_{1}}$ and $\vec{r_{2}}$ and center of mass distance $d=1.80$ nm); (b), (d) and (f) indicate three conformations during the \textit{circular} two-step manipulation (along two circular segments $\Gamma_1$ and $\Gamma_2$ and with distance $d=4.725$ nm).}
    \label{fig:Tip_manipulation}       % Give a unique label
\end{figure}

In a typical STM experiment, the torque cannot be applied to the gears directly. Instead, a handle gear is introduced as a mediator between STM tip and target gear\cite{WeiHyo2019}. To mimic this situation, we manipulate the handle gear along two specific trajectories as shown in Fig.\ \ref{fig:Tip_manipulation}, which will in turn drive the second gear counterclockwise.

For the molecule gears, we use a \textit{linear} two-step manipulation along two vectors $\bm{r}_1$ and $\bm{r}_2$ with a waiting time of $30$ ps between the two steps. For the solid-state gears we use a \textit{circular} two step manipulation path along the trajectories $\Gamma_1$ and $\Gamma_2$ due to large deformations occurring when using \textit{linear} paths.
Both manipulations are done with a fixed distance between the first and the second gear (before and after moving along the respective trajectory): for molecule gears we take $d=1.5$nm and for solid state gears $d=4.725$ nm. The distance between the second and the third gear is varied. From the perspective of the second gear, the first gear moves $60\degree $ per step, amounting to a total of $120\degree $.

\begin{figure}[t]
    \centering
    \includegraphics[width=\textwidth]{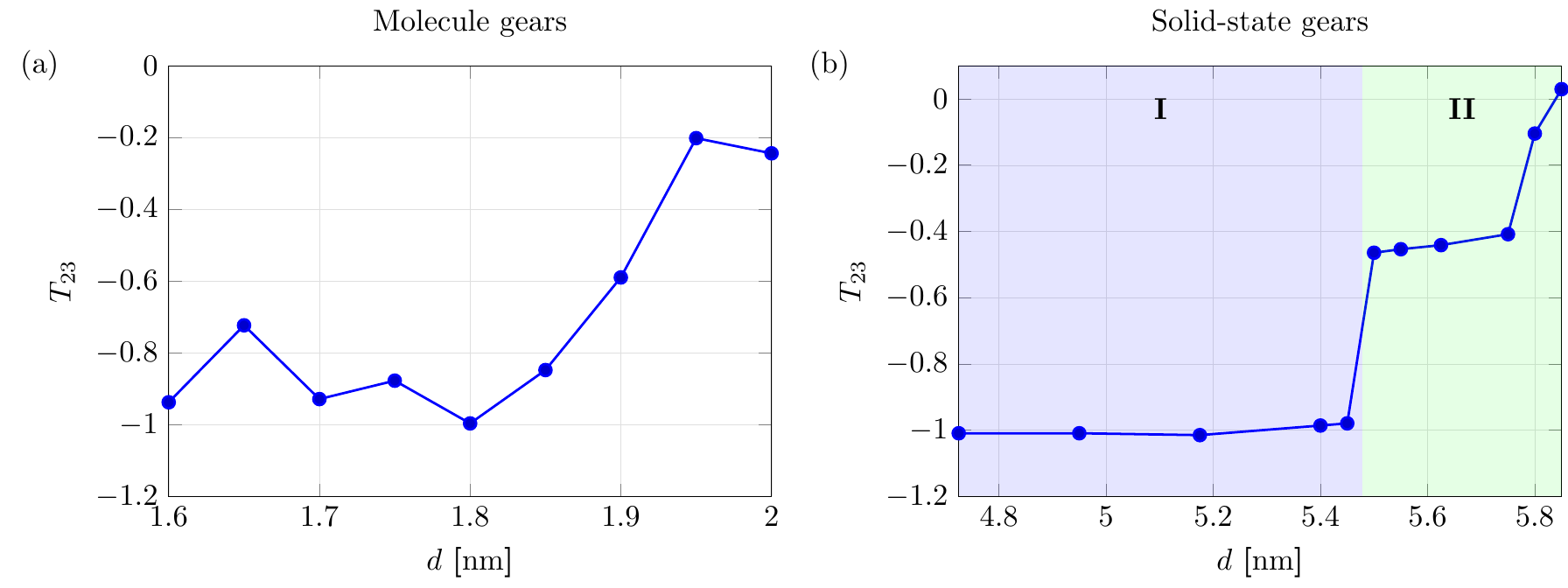}
    \caption{The transmission coefficient $T_{23}=\theta_3/\theta_{2}$ between the second and third gear for different center of mass distances $d$ with constant speed for (a) molecule gears during the \textit{linear} two-steps manipulation and (b) solid-state gears during the \textit{circular} two-steps manipulation, where $\theta_2$ and $\theta_3$ are the total angular displacements for the second(middle) and third(right) gear. For solid-state gears we can distinguish between region I (blue) and II (green) indicating the driving ($\mid T_{23}\mid\approx1$) and dragging ($0\leq\mid T_{23}\mid \leq 0.5$) regimes, respectively.}
    % (a) Shows a minimum at $d=1.8$nm with $T_{23}=-0.99$ and a drop for $d>1.8$nm (move to text)
    \label{fig:transmission}       % Give a unique label
\end{figure}

The results of the MD simulations are shown in Fig.\ \ref{fig:transmission}. For molecule gears, the movement took $1$ ns (excluding relaxation time) and covered a total distance of $3$ nm. For the solid-state gears it took $3.3$ ns and covered an arc length of $5$ nm. In order to compare the results, we define the transmission coefficient as follows:
\begin{equation}
    T_{23} = \frac{\theta_3}{\theta_2}\;,
\end{equation}
where $\theta_2$ and $\theta_3$ are the total angular displacements of the second and third gears, respectively. This quantity  describes  how well both gears are interlocked, even though the angular velocity cannot be obtained directly. For instance, {when the handle gear moves two circular-steps ($120\degree$ with respect to the second gear), the third gear will also rotate two steps in opposite direction}. One can use the NBRA to estimate the corresponding angle and to compute the transmission coefficient, which gives a value in the range $-1\leq T_{23}\leq 0$.

Figure \ref{fig:transmission} (a) shows the average transmission coefficient of $20$ simulations for different center of mass distances $d$ during the \textit{linear} two-step manipulation. For $d\leq1.9$ nm, we have similar interlocked rotations with $0.6\leq \mid T_{23}\mid\leqslant1$. The optimal collective rotation can be found at $d=1.8$ nm with $\mid T_{23}\mid\approx1$. For larger distances, we see a quick decay for transmission to around $\mid T_{23}\mid\approx0.2$.

In Fig.\ \ref{fig:transmission} (b), the average transmission coefficient of solid-state gears for different center of mass distance $d$ is shown. Here we can distinguish two different regions (highlighted in blue and green):

\paragraph{Region I:}
\begin{equation}
 \mid T_{23}\mid\approx 1
\end{equation}
For $d\leq 5.45$ nm, the gears are in \textit{driving phase}, with almost perfectly interlocked rotation.

\begin{figure}[t]
    \centering
    \includegraphics[width=\textwidth]{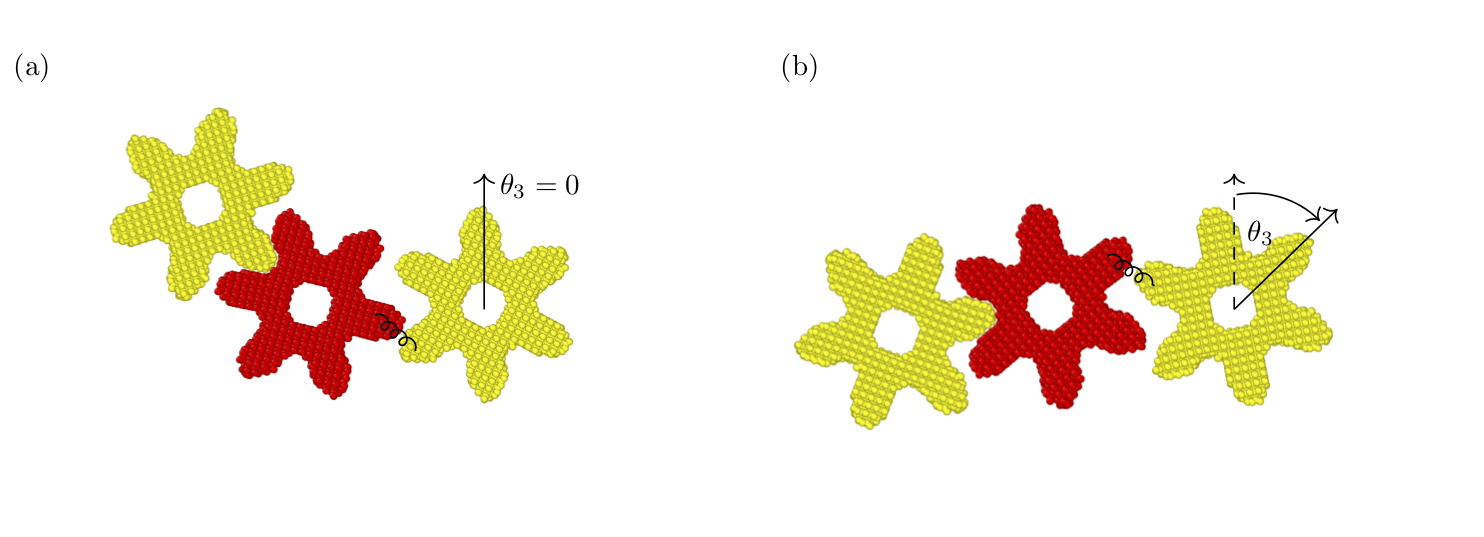}
    \caption{Schematic plots of the \textit{dragging phase}. (a) and (b) show two different conformations during the \textit{circular} two-step manipulation for center of mass distances $d\geq5.50$ (Region II). (a) While the first and the second gear rotates as shown in Fig.\ \ref{fig:Tip_manipulation}, the third gear does not interlock; it is still in its starting conformation ($\theta_3 = 0$). As the distances between the teeth become smaller, the middle gear starts to drag the right gear (represented by the string. (b) As the motion continues, the distance between the teeth becomes too big to sustain the drag and they lose contact. $\theta_3$ is the total angle covered by the motion of the right gear.}
    \label{fig:attracting_dragging}       % Give a unique label
\end{figure}

\paragraph{Region II:}
\begin{equation}
    0\leq\mid T_{23}\mid\leq 0.5
\end{equation}
For larger distances, we see a plateau between $5.50\leq d\leq5.75$ followed by another sudden decrease in $\mid  T_{23}\mid$. We call this region \textit{dragging phase}, the gears barely touch at their respective tips, and the rotation of the third gear is mainly driven by the attractive force between the atoms of the tips, as shown in Fig.\ \ref{fig:attracting_dragging} (a) and (b).

In Fig.\ \ref{fig:attracting_dragging} (a), while the first two gears undergo the first step in Fig.\ \ref{fig:Tip_manipulation}, the third does not interlock and stays in its starting conformation ($\theta_3=0$). When the distance between the teeth becomes sufficiently small, the middle gear starts to drag (highlighted by a spring) the right one (see Fig.\ \ref{fig:Tip_manipulation}). This motion will then continue until the distance between the teeth becomes too large to sustain the drag. The angle covered by the right gear due to the drag is $\theta_3$. In the end, this results in a decrease of $\mid T_{23}\mid$. For distances $d\geq5.85$ nm the gears are too far apart for any collective rotation to occur.

While there are two regions for the solid-state gears, the molecule gears do not show such a distinct pattern for changes in the center of mass distance. In comparison, their transmission coefficient is subject to much higher fluctuations for every change in the center of mass distance, whereas for solid-state gears significant changes only occur in the transition between the regions.

%%%%%%%%%%%%%%%%%%%%%%%%%%%%%%%%%%%%%%%%%%%%%%%%%%%%%%%%%%%%%%%%%%%%%%%%%%%%%%%%%%%%%%%%%%%%%%%%%%%%%%%%%%%%%%%%%%%%%%%
\section{Conclusions and Outlook}
\label{sec:conclusion}
In this chapter, we have carried out, using atomistic Molecular Dyanmics simulations, a comparative study of the transmission of rotational motion across molecule gears as well as solid-state gears. Our approach is based on a nearly rigid-body approximation, which helps to  define the orientation vector of the gear for weakly deformed structures. We discussed two possible strategies to induce a rotational motion of the leading gear: either by (i) applying an external torque or (ii) by mimicking the manipulation with an STM tip.  In the first case (i), the introduction of \textit{locking coefficients} allowed to clearly identify different rotational regimes, denoted as underdriving, driving and overdriving phases. It turns out that for molecule gears, collective rotations are possible only up to two gears, a result related to the dissipation of energy into internal molecular degrees of freedom. In contrast, the solid state gears largely preserve the rigid-body like character, so that collective rotations become possible  with up to ten gears.
Concerning case (ii), we found out that  transmission of rotational motion across more than two molecule gears is feasible and it critically depends on the center-of-mass distance between the gears. For for solid-state gears,  driving and dragging phases were identified, in dependence of the center of mass distance between the gears.

{Future computational studies will need to include the influence of a real substrate in order to address additional energy dissipation channels, which may hamper the efficient transmission of motion across a gear train. This problem is closely connected with the more general problem of the theoretical description of friction processes at the nanoscale{\cite{Panizon2018}}.  Elucidating the interaction mechanisms between nanoscale gears and various substrates builds an integral part of the understanding of the working principles of nanoscale machinery{\cite{Roegel2015}}.}
{Looking beyond the classical regime, the possibility of studying quantum effects in mechanical gears provides a fascinating perspective\cite{MacKinnon2002,Liu2019}.}

\begin{acknowledgement}
We would like to thank C. Joachim, A.~Kutscher, A.~Mendez, A.~Raptakis, T.~K{\"u}hne, D.~Bodesheim, S.~Kampmann, R.~Biele, D. Ryndyk, A.~Dianat, and F.~Moresco for very useful discussions and suggestions. This work has been supported by the International Max Planck Research School (IMPRS) for ``Many-Particle Systems in Structured Environments'' and also by the European Union Horizon 2020 FET Open project "Mechanics with Molecules" (MEMO, grant nr.\ 766864).
\end{acknowledgement}
%
%\bibliographystyle{spphys}
%\bibliography{reference}

\end{document}